\documentclass[a4paper]{article}

\usepackage{INTERSPEECH2022}

\title{Toward Zero Oracle Word Error Rate on the Switchboard Benchmark}
\name{Arlo Faria, Adam Janin, Korbinian Riedhammer, Sidhi Adkoli}
\address{Mod9 Technologies, Berkeley, CA, USA
}
\email{team@mod9.com}

\begin{document}

\maketitle

\begin{abstract}
The ``Switchboard benchmark'' is a very well-known test set in automatic speech recognition (ASR) research, establishing record-setting performance for systems that claim human-level transcription accuracy.
This work highlights lesser-known practical considerations of this evaluation, demonstrating major improvements in word error rate (WER) by correcting the reference transcriptions and deviating from the official scoring methodology.
In this more detailed and reproducible scheme, even commercial ASR systems can score below 5\% WER and the established record for a research system is lowered to 2.3\%.
An alternative metric of transcript precision is proposed, which does not penalize deletions and appears to be more discriminating for human vs. machine performance. 
While commercial ASR systems are still below this threshold, 
a research system is shown to clearly surpass the accuracy of commercial human speech recognition. 
This work also explores using standardized scoring tools to compute oracle WER by selecting the best among a list of alternatives.
A phrase alternatives representation is compared to utterance-level N-best lists and word-level data structures;
using dense lattices and adding out-of-vocabulary words, this achieves an oracle WER of 0.18\%.
\end{abstract}


\section{Introduction}

This work is about the very well-known ``Switchboard'' subset of an evaluation of US English conversational telephone speech recognition, originally conducted by NIST in 2000. \cite{fiscus20002000}

The current best published result is 4.3\% WER \cite{tuske2021limit}, which also acknowledged that
``most of the speakers appear in the training data, hyperparameters are optimized on [the test set], and the human error rate might also have been overestimated''.

Other recent results demonstrate 5.0\% with low-latency streaming \cite{nguyen2021super}, 
and many works reference \cite{XiongEtAl:msr-tr2016} and \cite{saon2017interspeech} as the first systems to achieve the milestone of parity with human performance, which is described as 5.1\% to 5.9\% WER.

A careful analysis in \cite{mansfield2021revisiting} notes that
``humans are more likely to miss words than to misrecognize them'',
and is notable in several regards: 
code was provided to specify a non-standard data cleaning and text normalization process,
while output from a research system was re-scored in an (unsuccessful) attempt to replicate a published result.
Our work continues in this effort to fully describe and improve upon the standard scoring methodology, 
sharing data and software to enable reproducible results.

This work benchmarks commercial ASR systems, inspired by \cite{del2021earnings}, 
which archived outputs from the dates of collection.
For this Switchboard benchmark, a particular advantage of benchmarking commercial systems is that the evaluation simulates a more realistic scenario of presenting each conversation side as an entire 5-minute audio file.
By contrast, the NIST evaluation allowed research systems to use the reference segmentation as input, 
which can result in artificially low WER scores.

This work is similar to \cite{kim2021semantic} by presenting transcript precision and recall as possibly more insightful alternatives to WER, particularly for highlighting characteristics of human performance.
The use of an ``oracle'' word error rate that is optimistically calculated from ASR alternatives is similar to \cite{moriya2021asr} which reranks N-best alternatives for spoken content retrieval, as well as our prior work \cite{wegmann2013tao} in evaluating systems for spoken term detection.

While evaluating traditional N-best lists, we also introduce a novel representation for phrase-level alternatives.  This captures the full expressiveness of an ASR lattice \cite{povey2012generating}, but in a more compact and linear data structure that can be conveniently manipulated as input to ASR scoring software, or indexed by a text-based search engine infrastructure.
The aim of this work is to show how this representation enables nearly perfect oracle accuracy (0.18\% WER) on a well-established ASR task.
This theoretical result motivates the further use of phrase alternatives toward a highly practical goal of enabling spoken term detection (i.e. audio search) applications that exhibit perfect recall.

\section{Scoring the Switchboard Benchmark}

\begin{table*}[t]
  \caption{
    Switchboard WER scored with corrected references, optional deletions \& exclusions, using differing segmentations.
    \\ Italicized results in all tables used the reference segmentation, 
    which can be considered a bound on expected real-world performance.
  }
  \label{tab:wer}
  \centering
  \begin{tabular}{ l r r r r r | r}
    \toprule
                                  & ASR1  & ASR2  & ASR3  & ASR4 & ASR5 & \textit{ASR6} \\
    \midrule
    \midrule
    LDC STM \& GLM                & 10.18 & 12.37 & 11.10 & 8.25 & 8.62 & \textit{4.63} \\
    + RT-03  GLM                  & 9.94  & 12.20 & 10.88 & 8.07 & 7.96 & \textit{4.40} \\
    + RT-04F GLM                  & 9.92  & 12.20 & 10.86 & 8.05 & 7.95 & \textit{4.39} \\
    \midrule
    \midrule
    STM with corrections          & 8.15  & 10.42 &  8.65 & 5.65 & 6.08 & \textit{2.72} \\
    + GLM with alternations       & 8.07  & 10.29 &  8.41 & 5.51 & 5.79 & \textit{2.63} \\
    \midrule
    + Exclude hesitations         & 7.86  &  9.99 &  8.41 & 5.28 & 5.30 & \textit{2.45} \\
    + Optional backchannels       & 7.77  &  9.77 &  7.55 & 5.17 & 4.54 & \textit{2.43} \\
    + Exclude backchannels        & 6.48  &  9.72 &  7.55 & 5.08 & 4.54 & \textit{2.37} \\
    \midrule
    + Single-segment STM          & \textbf{6.43} & \textbf{9.67} & \textbf{6.42} & \textbf{5.01} &\textbf{4.50} & \textit{2.30} \\
    + \textit{Reference segmentation} & \textit{5.94} & \textit{9.66} & \textit{5.09} & \textit{4.29} &\textit{4.03} & \textbf{\textit{2.30}} \\
    \bottomrule
  \end{tabular}
\end{table*}

\subsection{Corrected reference files}
Reference files from the original NIST evaluation are now distributed by the Linguistic Data Consortium (LDC)\footnote{https://catalog.ldc.upenn.edu/LDC2002T43},
but differ from what was later used in DARPA-funded evaluations known as ``RT-03'' and ``RT-04F''.
For example, the newer GLM files include backchannel mappings that generally improve scores.

Human transcribers disagree on this very difficult task
\cite{XiongEtAl:msr-tr2016,saon2017interspeech,mansfield2021revisiting,greenberg1996insights,stolcke2017comparing},
so it should not be surprising that there are inevitably some errors in these reference transcripts and mappings.
For this work, a professional linguist was commissioned to very carefully audit and correct these references.
In addition to the original transcripts, they could refer to four independent results from human speech recognition (HSR) services,
but not any of the ASR systems.
This paper's authors further corrected the GLM file with ad-hoc normalization of number formatting.

However, the vast majority of changes were related to an artifact of the \verb|make_reference| script 
that is distributed with the test set;
it was used to create the reference STM by converting transcripts from an original TXT file format.
Unfortunately, every contraction in the original transcript is always expanded into multiple words
(see lines 126-131 in \verb|make_reference|).
This does not seem to be sensible, 
especially considering that the GLM filtering would also redundantly expand all contractions.
We thus decided to reverse this automatic expansion of contractions,
and directed the highly skilled linguist to transcribe each instance correctly as either its contracted or expanded form 
by carefully listening to each acoustic realization,
favoring the contracted form in cases of true ambiguity.

These corrected reference files are shared publicly,\footnote{https://mod9.io/switchboard-benchmark.\{glm,stm\}}
and should lead to substantial improvements across all systems.

\subsection{Expansions vs. alternations}
The NIST SCTK software\footnote{https://github.com/usnistgov/SCTK} 
can use a GLM mapping file to filter reference STM and hypothesis CTM files by applying a set of transformation rules.
For example, contracted or compound words can be expanded with a rule such as
\verb|I'M => I AM|.

However, by always expanding contractions in both the reference STM and hypothesis CTM, 
this rule often double-counts correct matches as well as errors.
A better approach is to denote \textit{alternations} to be applied in the GLM file, e.g. 
\verb|I'M => { I'M / I AM }|, which will be scored as one or two matches or errors as appropriate.
The original form should be listed first in the alternation, since SCTK will favor it when multiple alignments have the same number of errors; otherwise, favoring the expanded form results in overly optimistic scoring.

\subsection{Optional deletions and excluded words}
Another effect of the filtering is to treat some words as \textit{optional deletions},
marked by parentheses in the STM file, in particular \texttt{(\%HESITATION)}.
An ASR system should exclude such difficult words from CTM hypotheses,
due to the asymmetric risk: 
an error can have a larger effect on the numerator of WER, compared to a correct match incrementing the denominator (Eq. \ref{eq:wer}).

One major commercial system (ASR3) never hypothesizes hesitations --
nor any backchannels such as ``uh-huh'', which are not optional deletions under the NIST scoring rules.
So that their system is not disadvantaged by a design choice, we can consider backchannels to be optional deletions as well.
So that other ASR systems are not then disadvantaged by hypothesizing backchannels, we also exclude those from their CTM files.

\subsection{Segmentation}
The NIST tools can misalign hypotheses with word-level timestamps that differ slightly from the reference utterance-level segmentation of an STM file.
A solution is to convert the multi-segment STM into one long segment.
This can improve WER for ASR systems with consistent timestamp drift, and is needed to score any HSR (human speech recognition) result.

This problem is not observed in academic research experiments,
because the reference segmentation is assumed to be a valid input to the ASR system.
This practice may be unrealistic in real-world settings, however, 
as seen in the bottom rows of \tablename~\ref{tab:wer}:
it can have a rather significant effect on WER results.

\subsection{Measuring accuracy with precision and recall}

\begin{equation}
  \label{eq:wer}
  \textrm{WER} =
  100\% \times
  \frac{\textrm{\#Inserted} + \textrm{\#Deleted} + \textrm{\#Substituted}}{\textrm{\#Correct} + \textrm{\#Deleted} + \textrm{\#Substituted}}
\end{equation}
\begin{equation}
  \label{eq:precision}
  \textrm{Precision} =
  \frac{\textrm{\#Correct}}{\textrm{\#Correct} + \textrm{\#Inserted} + \textrm{\#Substituted}}
\end{equation}
\begin{equation}
  \label{eq:recall}
  \textrm{Recall} =
  \frac{\textrm{\#Correct}}{\textrm{\#Correct} + \textrm{\#Deleted} + \textrm{\#Substituted}}
\end{equation}
Whereas the WER metric can be computed as in Eq. \ref{eq:wer},
a pair of non-standard metrics can also be useful to consider when evaluating ASR accuracy.
Transcript precision is the proportion of hypothesized words that are correct.
It does not penalize deletions and scores consistently well for human transcripts,
since it forgives the common tendency to omit words or phrases that do not convey much meaning 
(e.g. stuttering ``i i i i ...'').
The recall metric can be rather variable for HSR results;
it could be useful for evaluating against non-verbatim reference transcriptions.

\begin{table}[h]
  \caption{
    Automatic (ASR) vs. human (HSR) speech recognition.
    \\ Human speech recognition marked $^*$ was not speaker-labeled; it was scored against a force-aligned speaker-merged STM file.}
  \label{tab:precision}
  \centering
  \begin{tabular}{ l r c c r }
    \toprule
                  &           WER &     Precision &        Recall & Cost/min. \\
    \midrule
    \midrule
    ASR1          &          6.43 &         .950  &         .945  & --- \\
    ASR2          &          9.67 &         .930  &         .916  & 4.0\textcent{} \\
    ASR3          &          6.42 &         .953  &         .943  & 7.2\textcent{} \\
    ASR4          &          5.01 &         .961  &         .962  & 4.8\textcent{} \\
    ASR5          &          4.50 & \textbf{.964} &         .960  & 3.3\textcent{} \\
    \midrule
    \textit{ASR1} & \textit{5.94} & \textit{.953} & \textit{.947} & --- \\
    \textit{ASR2} & \textit{9.66} & \textit{.929} & \textit{.913} & 2.5\textcent{} \\
    \textit{ASR3} & \textit{5.09} & \textit{.965} & \textit{.953} & 16.5\textcent{} \\
    \textit{ASR4} & \textit{4.29} & \textit{.969} & \textit{.963} & 11.0\textcent{} \\
    \textit{ASR5} & \textit{4.01} & \textit{.968} & \textit{.964} & 3.0\textcent{} \\
    \textit{ASR6} & \textit{2.30} & \textbf{\textit{.981}} &\textit{.981} & --- \\
    \midrule
    \midrule
    HSR1          &           4.84 & \textbf{.973} &        .957  & \$1.25 \\
    HSR2          &           4.33 & \textbf{.975} &        .962  & \$2.75 \\
    \midrule
    HSR3$^*$      &          12.95 & \textbf{.973} &        .877  & \$0.79 \\
    HSR4$^*$      &          11.72 & \textbf{.972} &        .891  & \$2.00 \\
    \bottomrule
  \end{tabular}
\end{table}

\section{Representations of ASR Alternatives}

Lattices can be generated by some ASR decoders, particularly in a WFST system such as Kaldi \cite{povey2012generating},
to represent the inherent ambiguity and uncertainty of hypotheses.
However, the lattices are large and difficult to use in applications that
require properties such as time-synchronous word sub-sequences.

Let $L_u$ be the formal language representing the set of all word sequences encoded in the lattice for a given utterance $u$.

\subsection{Utterance-level alternatives (i.e. N-best lists)}
\begin{table}[t]
  \caption{
    Oracle WER for utterance-level ($N$-best) alternatives.
  }
  \label{tab:alternatives-utterance}
  \centering
  \begin{tabular}{ l c r r r r r }
    \toprule
                  &           WER &   $N$ & $N_{\max}$ & $N_{.9}$ & $N_{.5}$ &    MB \\
    \midrule
    \midrule
    \textit{ASR1} & \textit{4.61} &      2 &          2 &        2 &       2  &   0.2 \\
    \textit{ASR1} & \textit{2.70} &     10 &         10 &       10 &      10  &   0.5 \\
    \textit{ASR1} & \textit{1.58} &    100 &        100 &      100 &     100  &   1.9 \\
    \textit{ASR1} & \textit{1.09} &   1000 &       1000 &     1000 &    1000  &  15.2 \\
    \midrule
    \textit{ASR2} & \textit{7.39} &      2 &          2 &        2 &       2  &   0.2 \\
    \textit{ASR2} & \textit{5.41} &     10 &         10 &       10 &      10  &   0.5 \\
    \textit{ASR2} & \textit{4.35} &    100 &        100 &      100 &      29  &   1.5 \\
    \textit{ASR2} & \textit{4.05} &   1000 &       1000 &     1000 &      29  &   7.6 \\
    \midrule
    \textit{ASR3} & \textit{3.95} &      2 &          2 &        2 &       2  &   0.2 \\
    \textit{ASR3} & \textit{2.38} &     10 &         10 &       10 &       7  &   0.4 \\
    \textit{ASR3} & \textit{2.06} &$\infty$&         20 &       20 &       7  &   0.5 \\
    \midrule
    \textit{ASR4} & \textit{3.12} &      2 &          2 &        2 &       2  &   0.2 \\
    \textit{ASR4} & \textit{2.01} &$\infty$&         10 &       10 &      10  &   0.5 \\
    \midrule
    \textit{ASR5} & \textit{2.98} &      2 &          2 &        2 &       2  &   0.2 \\
    \textit{ASR5} & \textit{2.29} &$\infty$&          5 &        5 &       5  &   0.4 \\
    \bottomrule
  \end{tabular}
\end{table}

Utterance-level alternatives, better known as N-best lists, can be used to enumerate a formal language $L_u(N)$,
a set comprising up to $N$ most likely word sequences in the lattice.
The lattice's language is a superset, with equality in the theoretical limit:
\begin{equation}
L_u \supseteq \lim_{N \to \infty} L_u(N)
\end{equation}

\subsection{Word-level alternatives}

Word-level alternatives, sometimes known as \textit{sausages}, can be derived by aligning paths in a lattice \cite{mangu2000finding}
or from statistics used in Minimum Bayes' Risk decoding \cite{xu2011minimum}.
These represent a smaller formal language of up to $N$ single-word sequences $L_w(N)$ at each word position $w$.
Due to 1-to-1 word alignments, the lattice's language cannot be decomposed as a 
cross-product and concatenation (indicated by $\prod$) of component sets:
\begin{equation}
L_u \neq \prod_{w \in u} L_w(N)
\end{equation}
There may be sequences in $L_u$ that cannot be represented as a concatenation of elements in $L_w(N)$, even for large $N$.

\begin{table}[t]
  \caption{
    Oracle WER for word-level alternatives.
  }
  \label{tab:alternatives-word}
  \centering
  \begin{tabular}{ l c r r r r r }
    \toprule
                  &           WER &   $N$ & $N_{\max}$ & $N_{.9}$ & $N_{.5}$ &  MB \\
    \midrule
    \midrule
    \textit{ASR1} & \textit{2.69} &      2 &          2 &        2 &       2  & 0.2 \\
    \textit{ASR1} & \textit{1.35} &     10 &         10 &       10 &       2  & 0.4 \\
    \textit{ASR1} & \textit{1.19} &    100 &        100 &       12 &       2  & 0.5 \\
    \textit{ASR1} & \textit{1.19} &$\infty$&        323 &       12 &       2  & 0.5 \\
    \midrule
    \textit{ASR2} & \textit{6.98} &      2 &          2 &        2 &       1  & 0.2 \\
    \textit{ASR2} & \textit{5.75} &     10 &         10 &        3 &       1  & 0.2 \\
    \textit{ASR2} & \textit{5.74} &$\infty$&         25 &        3 &       1  & 0.2 \\
    \bottomrule
  \end{tabular}
\end{table}

\subsection{Phrase-level alternatives}

\begin{table}[t]
  \caption{Oracle WER for phrase-level alternatives.}
  \label{tab:alternatives-phrase}
  \centering
  \begin{tabular}{ l c r r r r r }
    \toprule
                  &           WER &   $N$ & $N_{\max}$ & $N_{.9}$ & $N_{.5}$ &  MB \\
    \midrule
    \midrule
    \textit{ASR1}     & \textit{2.92} &       2 &          2 &        2 &       2  & 0.3 \\
    \textit{ASR1}     & \textit{1.08} &      10 &         10 &       10 &       3  & 0.6 \\
    \textit{ASR1}     & \textit{0.65} &     100 &        100 &       22 &       3  & 1.0 \\
    \textit{ASR1}     & \textit{0.57} &    1000 &       1000 &       22 &       3  & 1.3 \\
    \bottomrule
  \end{tabular}
\end{table}

By contrast, all paths in the lattice can be represented as a subset of the crossed and concatenated phrase-level alternatives \cite{faria2021phrase}:
\begin{equation}
L_u \subseteq \lim_{N \to \infty} \prod_{p \in u} L_p(N) 
\end{equation}
In this formulation $L_p(N)$ is a set of up to $N$ word sequences, which may be of varying lengths, at phrase position $p$.

\subsection{Converting lattices to phrase alternatives}

Phrase alternatives can be derived from a lattice as follows:
\begin{enumerate}
\item Word-align the lattice, which may need determinization.
\item Establish phrase boundaries as those times not crossed by non-silence arcs (above some arc posterior threshold).
\item For each phrase, mask the lattice arcs outside the phrase boundaries by setting their output symbols as epsilon.
\item Determinize each phrase-masked lattice, which removes most epsilon arcs, and find $N$ best paths (i.e. phrases).
\end{enumerate}

The phrase alternatives representation is motivated by its compactness compared to utterance-level alternatives,
since it decomposes the utterance as a concatenation of word sequences that are assumed to be independent of each other.
It is also more expressive since this cross product generates additional word sequences that may not have been present in the lattice.

\subsection{Representing alternative hypotheses in NIST SCTK}
A lesser-known feature of the CTM file format is that it can be used to represent \textit{alternatives} in ASR hypotheses, for example:
\begin{verbatim}
sw_4390 A * * <ALT_BEGIN>
sw_4390 A 4.49 0.66 UM
sw_4390 A * * <ALT>
sw_4390 A 4.49 0.66 I'M
sw_4390 A * * <ALT_END>
\end{verbatim}
While this is typically used to represent \textit{alternations} created by filtering with the GLM file,
it can be further leveraged to enable oracle scoring of ASR alternatives at various levels.
However, this functionality requires a minor modification\footnote{https://github.com/usnistgov/SCTK/pull/34} to the \verb|sclite| source code,
as well as auxiliary software\footnote{https://pypi.org/project/mod9-asr} that can create the CTM files while fixing a couple of related bugs in SCTK
(such as expanding doubly-nested alternatives after GLM filtering).

\section{Speech Recognition Systems}
Automatic (ASR) and human (HSR) systems were evaluated:

\

\textbf{ASR1} is a Kaldi baseline. An OPGRU acoustic model and a trigram language model were trained only on Switchboard plus Fisher.  These models were loaded by the Mod9 ASR Engine to produce utterance-, word-, and phrase-level alternatives.

\textbf{ASR1$^*$} customized the decoding graph by adding the 28 words that were out-of-vocabulary (OOV)
with respect to the system's relatively small lexicon (about 40,000 words that appeared in the training data).
Pronunciations were automatically generated with a grapheme-to-phoneme model \cite{novak2016phonetisaurus}
by requesting the Mod9 ASR Engine's \verb|add-words| command. 
    
\textbf{ASR1$^\dagger$} used non-default pruning beam sizes to produce denser lattices, by requesting a \verb|speed:3| option of the Mod9 ASR Engine, a trade-off with more compute and memory usage.

\textbf{ASR1$^{*\dagger}$} combined both of the above settings.

\

\textbf{ASR2} is IBM Watson with an older ``Narrowband'' model,  
instead of using a more accurate ``next-generation'' model, 
because this system is uniquely capable of demonstrating utterance- and word-level alternatives at extreme depths.

\textbf{ASR3} is Google Cloud Platform's STT service, using an ``enhanced'' variant of their ``phone\_call'' model.
%%Their terms allow benchmarking, but publication requires written permission [currently pending].

\textbf{ASR4} is Amazon Transcribe, configured for US English.
%%Their terms allow benchmarking, if reproducible and reciprocal.

\textbf{ASR5} is Microsoft Azure's Speech-to-Text service, which generates utterance-level alternatives of very limited depth.

\

\textbf{ASR6} is the system in \cite{tuske2021limit}, from which IBM Research shared CTM-formatted system outputs for evaluation purposes.

\

\textbf{HSR1} is the Rev.com service, which has speaker labeling.

\textbf{HSR2} is the TranscribeMe service, requesting ``verbatim'' quality transcripts that include speaker labeling.

\textbf{HSR3} is the TranscribeMe service, requesting ``first draft'' quality transcripts that do not include speaker labeling.

\textbf{HSR4} is the cielo24 service, with no speaker labeling.

\section{Results}
All results can be reproduced from system outputs\footnote{https://mod9.io/switchboard-benchmark-results.tar.gz}
that were archived in early 2022,
using open-source scoring scripts.\footnote{https://mod9.io/switchboard-benchmark-scripts.tar.gz}

The bottom row and right column of \tablename~\ref{tab:wer}, middle section of \tablename~\ref{tab:precision}, and left columns of other tables have italicized font.
This convention is used to clarify which results might be considered unrealistic,
due to use of a reference segmentation or also because of the oracle nature of selecting a best alternative.

\tablename~\ref{tab:wer} presents the WER results from scoring each of the ASR systems with successively improved configurations of the scoring tools,
as described in Sections 2.1 through 2.4.

\tablename~\ref{tab:precision} compares the ASR and HSR systems, including precision and recall metrics in addition to WER.
The results for HSR3 and HSR4 are exceptional because they required conversion of reference STM files into a single-channel format,
using forced-alignment with an HTK-based ASR system; regions of overlapped speech may be incorrectly merged in some cases.
Dual-channel audio files were submitted to the HSR services, so transcribers could understand conversations sides in context.

\tablename~\ref{tab:precision} also reports the cost of processing the Switchboard test set, based on its duration of 100 minutes.
For ASR without reference segmentation, audio was presented as channel-separated files, thus totaling 200 minutes, much of which was silence.
For ASR that exploited reference segmentation, audio was presented as a collection of 1,834 short audio files, totaling 123 minutes.
Note: ASR3 and ASR4 costs increase even as less data is processed, since their respective policies are to bill requests by rounding up to 15s granularity or at minimum 15s.

Tables 3, 4, 5, and 6 report the oracle WER when the NIST SCTK scoring software is presented with CTM files that represent utterance-, word-, and phrase-level alternatives.
These results all use the reference segmentation, since the software cannot score alternatives that cross STM segment boundaries.
Each table reports the parameter $N$ that was requested, which may be greater than the actual $N_{\max}$ returned.
The $N_{.9}$ and $N_{.5}$ columns indicate the depths of alternatives at the top decile and median results; these convey the distribution more clearly than the mean statistic.
The rightmost columns report the storage size of the \verb|gzip|-compressed CTM files in megabytes.

The last row of Table \ref{tab:alternatives-phrase-2} relates a hypothetical oracle selecting the best transcript from a phrase-level representation of alternatives, derived from very dense lattices, decoded with added knowledge of all OOV words, using a reference segmentation.

\begin{table}[t]
  \caption{
    Oracle WER for phrase-level alternatives: adding all OOV words (ASR1$^*$); 
    denser lattices (ASR$^\dagger$); and both (ASR$^{*\dagger}$).
  }
  \label{tab:alternatives-phrase-2}
  \centering
  \begin{tabular}{ l c r r r r r }
    \toprule
                               &         WER   &      $N$ & $N_{\max}$ & $N_{.9}$ & $N_{.5}$ &  MB \\
    \midrule
    \midrule
    \textit{ASR1}$^*$          & \textit{5.79} &        1 &          1 &        1 &       1  & 0.1 \\
    \textit{ASR1}$^*$          & \textit{0.49} &      100 &        100 &       22 &       3  & 1.0 \\
    \textit{ASR1}$^*$          & \textit{0.42} & $\infty$ &       5250 &       22 &       3  & 1.4 \\
    \midrule
    \textit{ASR1}$^\dagger$    & \textit{0.36} &     1000 &       1000 &      125 &      14  & 5.4 \\
    \textit{ASR1}$^\dagger$    & \textit{0.33} &    10000 &      10000 &      125 &      14  & 7.6 \\
    \midrule
    \textit{ASR1}$^{*\dagger}$ & \textit{0.21} &     1000 &       1000 &      124 &      14  & 5.4 \\
    \textit{ASR1}$^{*\dagger}$ & \textbf{\textit{0.18}} &    10000 &      10000 &      124 &      14  & 7.4 \\
    \bottomrule
  \end{tabular}
\end{table}

\section{Conclusion}
This work highlighted subtle issues with evaluating the famous Switchboard benchmark.
It presented a reproducible Kaldi ASR baseline, comparing major cloud platforms to 
human transcription services, and clarified that IBM's research system achieves a \textbf{super-human record of 2.3\% instead of 4.3\% WER}.

Some experiments are unrealistic to varying degrees, ranging from the assumption of an oracle to the accepted use of a reference segmentation.
Nonetheless, such results demonstrate the potential for \textbf{lattice-based ASR approaching 0.18\% WER}.

These results motivate future work to improve lattice generation \cite{rybach2017lattice,lv2021let}, 
particularly in E2E ASR systems.
Our current research also explores open-vocabulary decoding in a WFST framework, 
in which novel words may be included in a lattice and derived phrase alternatives.
These advances enable new applications, e.g. audio search or machine-assisted transcription,
that can be designed to mitigate inevitable errors in 1-best ASR.

\section{Acknowledgments}
Thanks to our many friends from ICSI: \\
$\bigstar$ Michael Ellsworth, who carefully audited the references.\\
$\bigstar$ Andreas Stolcke, who clarified many evaluation practices.\\
$\bigstar$ Brian Kingsbury, who shared results from IBM Research.\\
$\bigstar$ Deanna Gelbart, who wrote code for phrase alternatives.

\vfill

\bibliographystyle{IEEEtran}
\bibliography{mybib}

\end{document}